\documentclass[manuscript]{aastex}

\usepackage{graphicx}
\usepackage{float}

\shorttitle{Runaway Dwarf Carbon Stars}
\shortauthors{Plant et al.}

\begin{document}

\title{Runaway Dwarf Carbon Stars as Candidate Supernova Ejecta}

\author{Kathryn A. Plant, Bruce Margon, Puragra Guhathakurta, Emily C. Cunningham, and Elisa Toloba\altaffilmark{1}}
\affil{Department of Astronomy \& Astrophysics and University of California Observatories, University of California, Santa Cruz, 1156 High Street, Santa Cruz, CA 95064}

\altaffiltext{1}{Current address: Texas Tech University, Physics Department, Box 41051, Lubbock, TX 79409-1051}
\email{kaplant@ucsc.edu}

\and

\author{Jeffrey A. Munn}
\affil{ US Naval Observatory, Flagstaff Station, 10391 West Naval Observatory Road, Flagstaff, AZ 86005-8521} 

\begin{abstract}

  The dwarf carbon (dC) star SDSS J112801.67+004034.6 has an unusually high radial velocity, 531$\pm 4$~km~s$^{-1}$.  We present proper motion and new spectroscopic observations which imply a large Galactic rest frame velocity, 425$\pm 9$~km~s$^{-1}$. 
 Several other SDSS dC stars are also inferred to have very high galactocentric velocities, again each based on both high heliocentric radial velocity and also confidently detected proper motions. Extreme velocities and the presence of $C_2$ bands in the spectra of dwarf stars are both rare.  Passage near the Galactic center can accelerate stars to such extreme velocities, but the large orbital angular momentum of SDSS J1128 precludes this explanation. Ejection from a supernova in a binary system or disruption of  a binary by other stars are possibilities, particularly as dC stars are thought to obtain their photospheric $C_2$ via mass transfer from an evolved companion.
\end{abstract}

 
 \clearpage

\section{Introduction}
 The presence of molecular carbon absorption bands is remarkable in the spectra of stars which have not yet evolved through the helium-burning and dredge-up phases that deposit carbon in the photosphere. The Sloan Digital Sky Survey (SDSS, York et al. 2000) has resulted in the discovery of a large number of faint high latitude carbon stars (Margon et al. 2002, Downes et al. 2004, Green 2013).  Several hundred of these stars show significant proper motion, and thus are most likely to be relatively nearby cool dwarfs rather than distant giants.  At this stage of evolution, these stars cannot have produced the carbon in their photospheres. Thus, dwarf carbon (dC) stars are generally considered members of post-mass transfer binaries, with the dwarf star polluted by an evolved, almost always undetected, companion.\\
\indent
 During a study of the kinematics of dC stars in an attempt to clarify
their population and evolution, we have noted that the distribution of radial velocities of stars in the \citet{gre13} SDSS dC sample has a few stars with extreme velocities, approaching Galactic escape velocity. As a probe for unusual stellar acceleration mechanisms, extreme kinematics are particularly interesting in a group of stars which likely were at one time members of close binaries.   
Most extreme stellar acceleration is caused by the disruption of a binary system either in an interaction with another star(s) (Poveda et al. 1967), or alternatively when the supernova of one member ejects the other star at high velocity (Zwicky 1957, Blaauw 1961).  Two special cases of the former involve passage near the Galactic center (Hills 1988, Manukian et al. 2013), and most observed hypervelocity stars escaping the galaxy  are O and B type stars (e.g. Brown 2015), whose trajectories are found to be consistent with these scenarios.
 \section{Sample Selection, Astrometry, and Spectroscopy}
\begin{figure}[H]
\includegraphics[width = 1.0\textwidth]{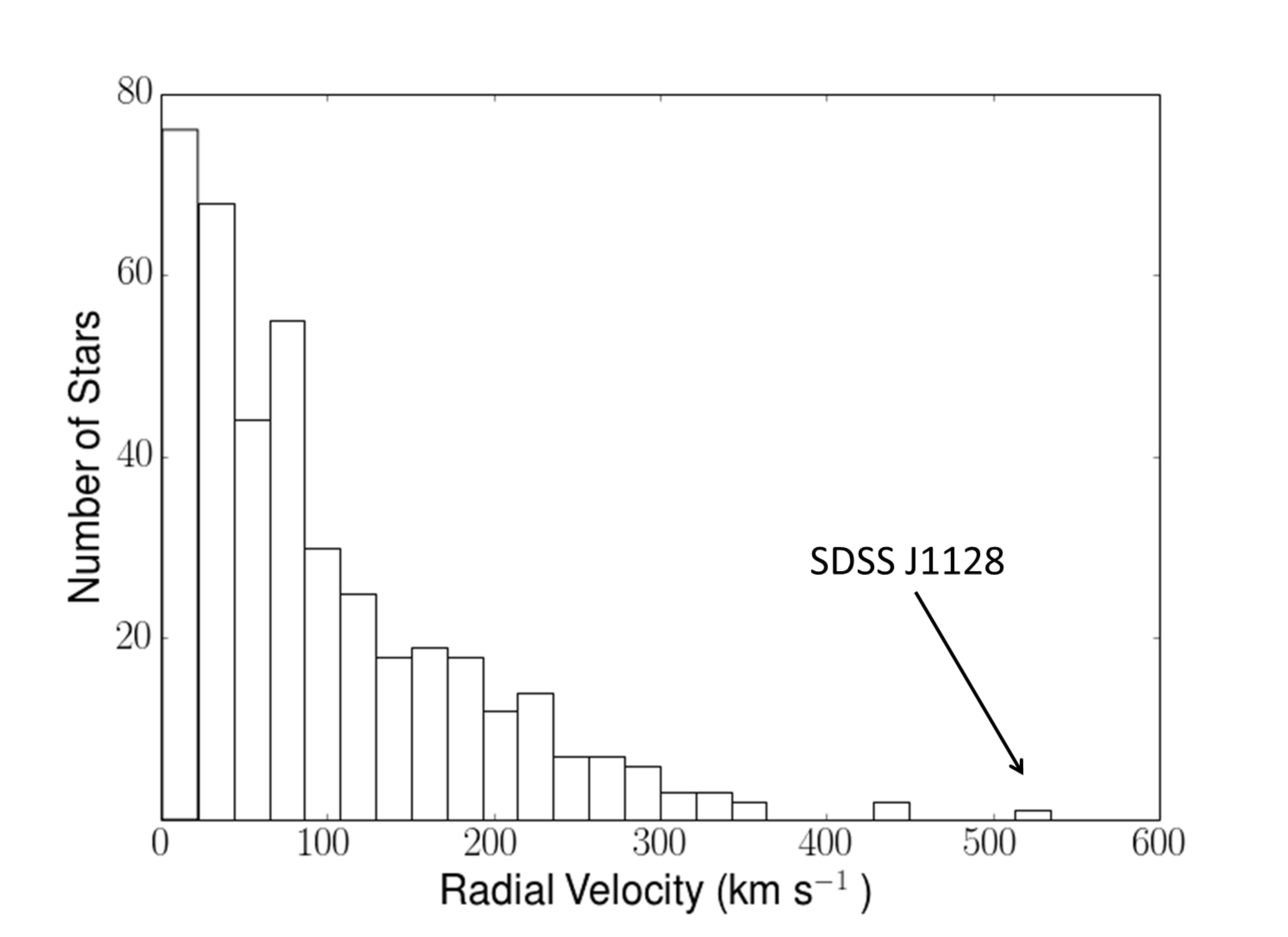}
\caption{Absolute value of the SDSS-tabulated heliocentric radial velocities of 410 dC stars. Note the three stars exceeding 
400 km  s$^{-1} $ .}
\label{velocityhistogram}
\end{figure}

We have searched for extreme runaway dC stars by selecting the stars with the largest radial velocities from a subset of stars from the \citet{gre13} SDSS dC catalog.
We formed the subset  by selecting stars with significant proper motions in the catalog of \citet{mun04, mun08}, which derives proper motions from combined SDSS and USNO-B \citep{mon03} astrometry, yielding a sample of several hundred dC stars.  Following the \cite{mun04} quality criteria (see also Kilic et al. 2006), we filter for stars that have detections in at least five survey epochs, have rms residuals less than 350 mas in each coordinate in the proper motion fit, have proper motions $3\times$ larger than its uncertainty in right ascension or declination, and have a one-to-one match between SDSS and USNO-B.  The significant proper motions are required for confident classification of the objects as dwarfs rather than distant giants. \\
\indent
We identify runaways by selecting the dC stars with the highest SDSS radial velocities.  Throughout this paper, we employ SDSS radial velocities measured by the SEGUE Stellar Parameter Pipeline (Lee et al. 2008), which calculates velocity using the Elodie template library of stellar spectra \citep{pru01}. These velocities may differ somewhat from those found for the same objects in older SDSS publications and online data portals.  For the objects discussed here, the newer SEGUE pipeline is preferred, as the velocity templates include C stars. Figure \ref{velocityhistogram} shows the velocity distribution for our sample. The notable runaway dC stars are the extreme radial velocity outliers within this sample. With an SDSS radial velocity of $550$~km~s$^{-1}$, the $r\sim19$ star SDSS J112801.67+004034.6 (hereafter J1128) is the most extreme of these objects, and this high radial velocity has been previously noted by \citet{mar02}.\\
\indent
Four extreme runaway candidates lie within the recent deep proper motion survey of Munn et al. (2014), which derives proper motions by combining SDSS astrometry with astrometry from later epoch CCD observations on the Steward Bok 2.3m and U.S Naval Observatory 1.3m telescopes, and which thus is both deeper than and independent of USNO-B.  For J1128, the proper motion in this survey is $\mu_{\alpha}$cos$(\delta)=-3.0 ~$mas~yr~$^{-1}$, $\mu_{\delta}=-24.3~$mas~yr$^{-1}$, with uncertainty of $2~$mas~yr$^{-1}$ in each coordinate, consistent with, but more precise than, the proper motion in the Munn et al. 2004 catalog.  Further, the SDSS, USNO-B, and USNO 1.3m images were visually  inspected to verify the reliability of the proper motion of J1128, thus confirming J1128 to be a relatively nearby dwarf, not a distant giant.\\
 \indent As the canonical model for the origin of dC stars requires the presence of a binary companion at least at some point in the system's lifetime, it is important to confirm that the high radial velocity is not in large part due to a binary orbit. 
We obtained spectra of J1128 using the Keck II telescope and DEIMOS spectrograph, in order to check for radial velocity variations. Spectra were obtained on 13 April 2015 and 7 April 2016 with the 600 line~mm$^{-1}$ grating with a central wavelength of 7200 \AA~ and resolution of R~=~2000 at H$\alpha$. The spectra were processed by a modified version of the spec2d pipeline developed by the DEEP2 team at UC Berkeley (Cooper et al. 2012 as modified by Kirby et al. 2015). Radial velocities are calculated using the Penalized Pixel Fitting method of \citet{cap04}. The 2015 spectrum yields a heliocentric radial velocity of $548~\pm$~7~ km~s$^{-1}$, with the uncertainties calculated as described by Toloba et al. (2016).  This result is close to the published SDSS value ($535~\pm$~6~ km~s$^{-1}$). The Keck and SDSS spectra are separated by 15 years.\\
\indent
The 2016 Keck spectrum, which is of lowest quality of any during the observing run, yields a discrepant value, $478~\pm~11$~km~s$^{-1}$. 
 Further observations are required, and we do not claim this as evidence of detection of binarity. Even if orbital motion has been detected, it is likely a small fraction of the total measured velocity, and thus the majority of the large radial velocity remains systemic. In the following analysis, we use the weighted mean of all the measured velocities ($531~\pm~4$~km~s$^{-1}$) as an estimate of the center-of-mass velocity. \\
Direct observational evidence for the binary nature of most dC stars remains elusive. Keck velocities of the other high velocity dC stars observed during the same observing run with the same instrument configuration show good agreement among repeat observations, and with the corresponding SDSS values. 
\section{Motion in the Galactic Frame}
\begin{figure}
\includegraphics[width=1.0 \textwidth]{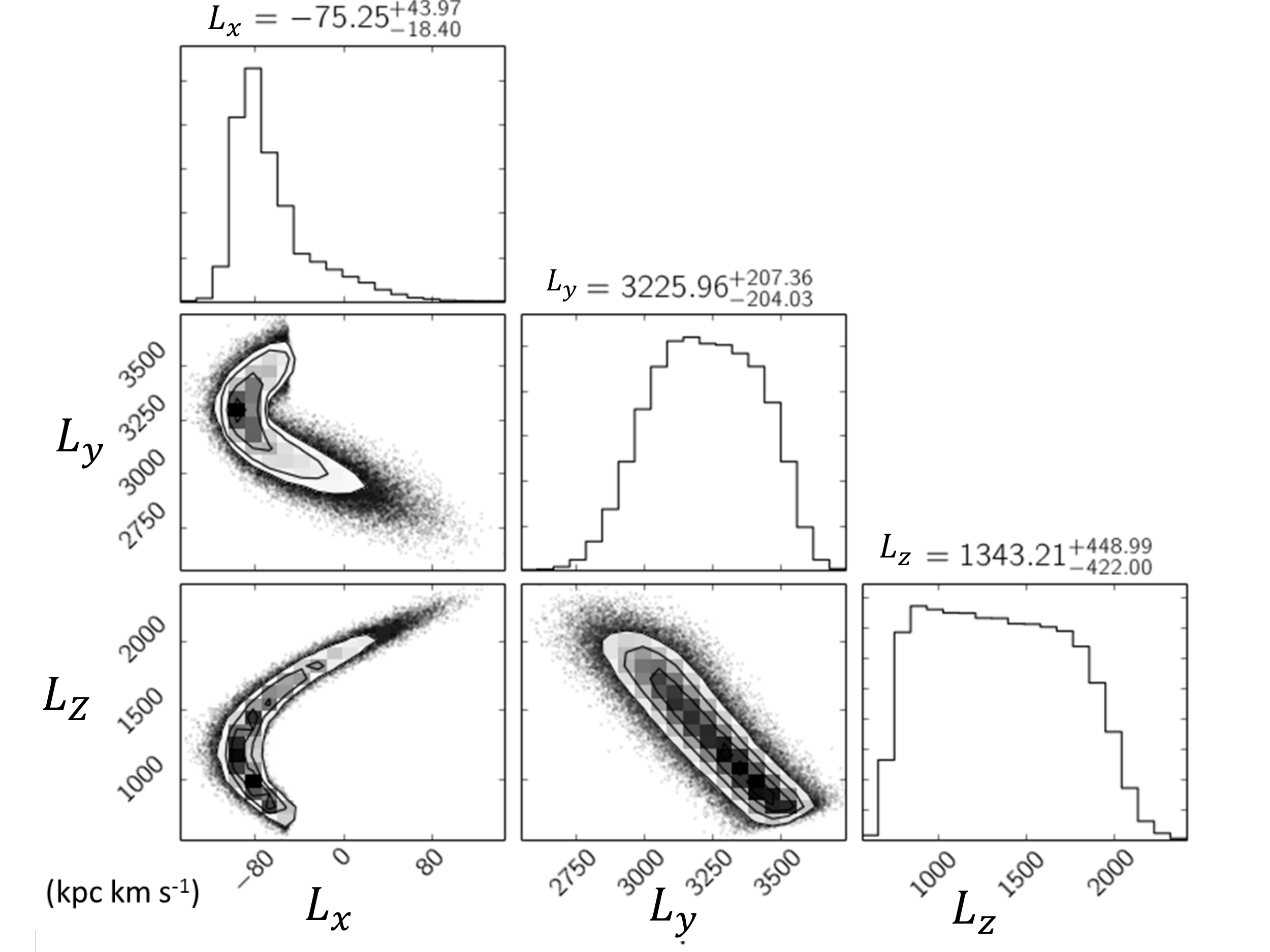}
\caption{Monte Carlo simulation for J1128 for specific angular momentum about the Galactic center. X, Y, and Z refer to the components of the angular momentum in Galactic cartesian coordinates. \textit{Contour plots:} $10^5$ Monte Carlo trials projected into 2 dimensions. \textit{Histograms:} Projected distributions for each component, as counts as a function of kpc km s$^{-1}$.  The significant specific angular momentum of J1128 excludes the Galactic center as an origin. The titles above each component distribution show the median value of the distribution and the 68\% confidence interval. Note that these order statistics are different from (and more robust than) the standard deviation since the distributions are not Gaussian. }
\label{cornerL}
\end{figure}

\begin{figure}
\centering
\includegraphics[width=1 \textwidth]{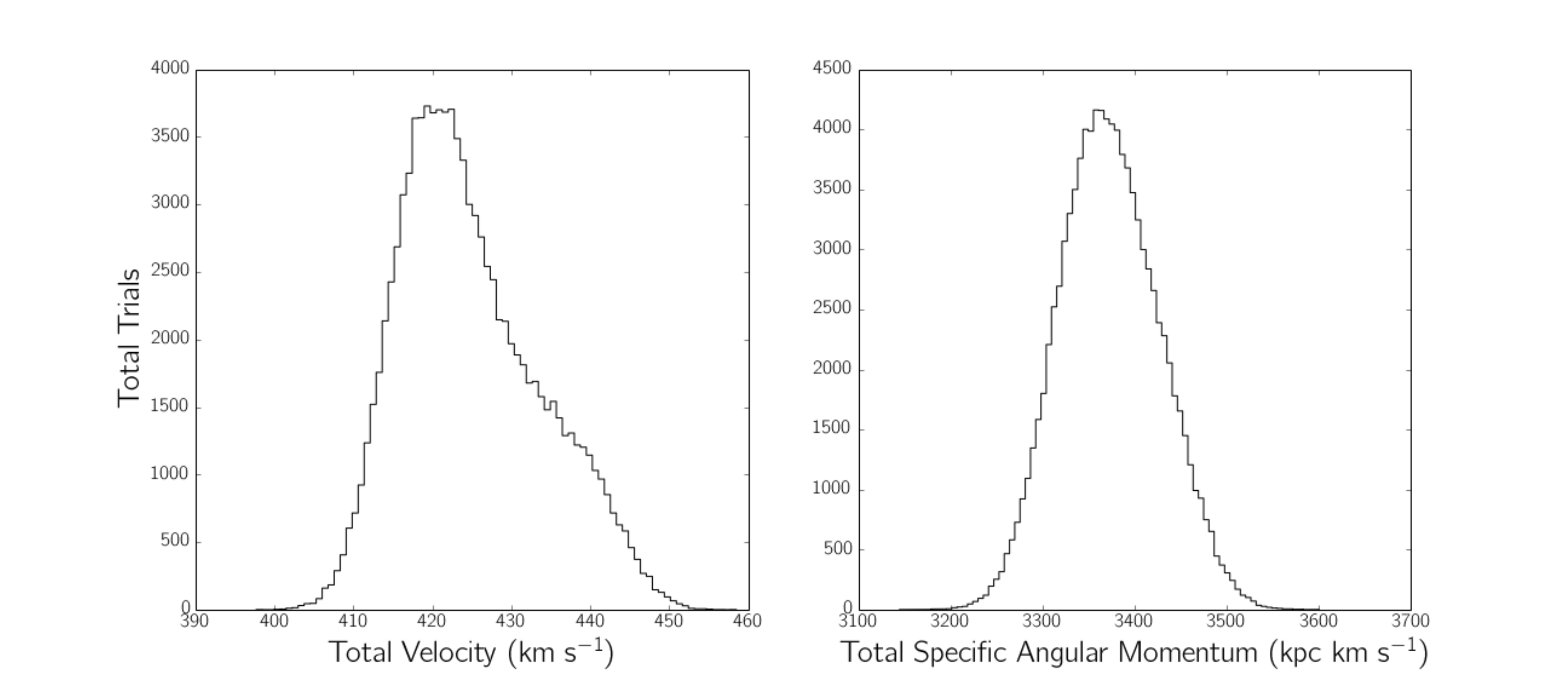}
\caption{Total motion of J1128, as derived by $10^5$ Monte Carlo trials.  \textit{Left panel:} Distribution of total velocity in the Galactic rest frame.  \textit{Right panel:} Distribution of total specific angular momentum about the center of the Galaxy.  Note that despite the uncertainties in the individual parameters, the total velocity is reasonably well-constrained and the large specific angular momentum precludes a Galactic center passage.}

\label{totals}
\end{figure}

\indent The 531~km s$^{-1}$ radial velocity and significant proper motion suggest that an unusual process accelerated J1128 to a velocity approaching Galactic escape velocity. Investigation of the possible acceleration mechanisms and the dynamic history of J1128 requires knowledge of the star's trajectory with respect to the Galactic center. The distance to J1128 is the least well-constrained of the parameters needed in estimating its motion in the Galactic rest frame. Although a traditional approach to estimate the distance would involve modeling atmospheric parameters to derive a spectroscopic distance, we are aware of no self-consistent model atmospheres currently available for dC stars. Instead we constrain this distance via reference to known dC luminosities. The US Naval Observatory has been conducting a parallax program for dC stars for many years (Harris et al. 1998, Harris 2016), and luminosities are now available for about 20 objects, with a range  $7.5 < M_r < 11.6$.  As a very conservative estimate for J1128, we simply assume that its luminosity lies somewhere within this interval, indicating possible distances in the range of about $0.3 < d < 2~$kpc.\\
\indent
To transform the heliocentric motion to the Galactic reference frame, we obtain the distributions of the kinematic parameters in the Galactic frame by Monte Carlo simulation (Figure \ref{cornerL}). We convert heliocentric radial velocity, proper motion, and distance to a right-handed Cartesian coordinate system defined with the Galactic Center at rest at the origin. The Z axis is perpendicular to the plane of the galaxy, the X axis points away from the Sun, and the Y axis points in the direction of the Sun's orbit.  We use a solar position X~=~$-7.8$~kpc, Y~=~0, Z~=~0 and solar velocity $v_X$ =~11.1~km~s $^{-1}$, $v_Y$~=~232.2~km~s$^{-1}$, $v_Z$~=~7.3~km~s$^{-1}$ \citep{zho14}.  Distributions were calculated by sampling the heliocentric parameter space (radial velocity, proper motion, and distance) according to the uncertainties in each parameter. Gaussian distributions are used for the parameters with well-constrained uncertainties-- proper motion and radial velocity. A flat distribution is used for distance, within the bounds set by the estimated luminosity (limits of 0.3 -- 2.0 kpc for heliocentric distance).\\
\indent The result of $10^5$ samplings yields a total galactocentric velocity with mean and standard deviation of $425\pm9$~km~s$^{-1}$. 
The motion is mostly away from the disk, ($v_z = 390 ^{+31} _{-32}$ km~s$^{-1}$) and opposing the direction of disk rotation ($v_y = -158 ^{+51} _{-52}$ km~s$^{-1}$).  Very little motion is in the X direction ($v_x = 16 ^{+27} _{-22}$ km~s$^{-1}$).  Due to the non-Gaussian distributions in the Galactic frame, uncertainties are given by order statistics, as the 68\% confidence interval. \\
\indent
Specific angular momentum, defined as $\textbf{ L} = \textbf{r} \times \textbf{v} $, provides useful information about possible trajectories.  Figure \ref{cornerL} shows the momentum results of the Monte Carlo simulation. Note that some components of the motion are more sensitive than others to the uncertainty in the distance, evident in the flattened peaks in the distributions for those components in Figure \ref{cornerL}. $L_z$ and $L_y$ are anticorrelated because $v_x$ is much smaller than the other velocity components and so the terms $L_z = x v_y - y v_x$ and $L_y = z v_x - x v_z $ are both dominated by the X position, which is sensitive to the uncertain distance. A large angular momentum forbids a primarily radial trajectory away from the coordinate origin. As the Z-distance of J1128 is small compared to its galactocentric distance, and  most of its motion is upwards out of the disk, the large specific angular momentum about the Galactic center excludes the Galactic center as an origin.  The results for the total velocity and and total specific angular momentum are summarized in Figure \ref{totals}.\\
\indent A preliminary analysis of several other dC stars in the high velocity tail of the distribution of Figure~1 yields additional candidates with high galactocentric velocities, although not as extreme as J1128, and preliminary trajectories also incompatible with a Galactic center passage. These results will be presented elsewhere. \\
\section{Discussion and Conclusions}
A large velocity for J1128 has been measured in both the radial and transverse directions. 
With a total galactocentric velocity of  $425 \pm 9~$km s$^{-1}$, J1128 is bound to the Galaxy. The large specific angular momentum  excludes a passage through the Galactic center as the acceleration mechanism. The extreme motion, together with the likely requirement that the object received its photospheric carbon from an evolved companion, suggests ejection from a binary system.  \citet{tau15} has shown that such high velocities are theoretically possible in supernova ejections and \cite{gva09} and others have shown that dynamical ejection (typically occurring in dense stellar clusters) can also produce this velocity, although for both these scenarios, few of the ejected stars have velocities exceeding 200 km s$^{-1}$ (\citet{por00}, \citet{per12}). \\ 
\indent
Historically, most extreme stellar acceleration was discussed in the context of O and B stars (e.g. Blaauw 1961, Poveda et al 1967, Hills 1988, Blaauw 1993, Leonard \& Duncan 1990, Portegies Zwart 2000, Hoogerwerf et al. 2001, Silva \& Napiwotzki 2011).  Now, large stellar datasets from Sloan Digital Sky Survey and LAMOST have increased the discovery rate of cooler, less massive high-velocity stars (e.g., Li et al. 2012, Favia et al. 2015, Zhong et al. 2014, Vickers et al. 2015).  A few of these stars approach or exceed Galactic escape velocity and, along with two recently discovered  O/B type hot-subdwarfs (Geier et al. 2015; N\'emeth et. al. (2016)), cannot have transited the Galactic center.  
J1128 thus joins the small group of late type stars suggested as high velocity ejecta from binaries. \\
\indent
Binary ejection predicts higher velocities for lower mass runaways, in both the supernova ejection (Portegies Zwart 2000) and ejection via dynamic disruption (Leonard \& Duncan 1990) mechanisms.   Portegies Zwart (2000) has predicted that many runaways accelerated by supernova remain bound to their high-speed neutron star companion, but that due to the lifetimes of late type stars, the pulsar would most likely have become unobservable.  In a dynamical ejection scenario, however, runaways exceeding 200 km~s~$^{-1}$ are not expected to remain bound to a companion \citep{per12}.  Thus, radial velocity monitoring continues to be interesting for J1128. Both types of ejection scenarios are most likely to occur in clusters or star forming regions, but tracing a runaway star to its birthplace is much more difficult for a long-lived K dwarf than than for the O/B runaways.  Given the carbon absorption bands in the current spectrum, mass transfer simulations could help constrain the dynamics of the original binary in order to provide evidence in favor of one mechanism or the other. \\
\indent

One possible clue may be the rarity of both dC stars and of high velocity ejections. Very high velocity dC stars are  $\sim10^{-2}$ of the total number of known dC stars. Dynamic ejection from a binary should presumably not favor the extremely rare dC star over other low mass stars in binaries, leading to the expectation of far more very high velocity non-carbon  runaway stars than dC stars of similar mass and velocity.  This is not observed. For some runaways \citep{gva09single}, dynamic ejection is favored in part because the disrupted binary was too young for the companion to have evolved to end in a supernova. For dwarf carbon stars, the carbon bands strongly suggest that the ejection event occurred very late in the evolution of the companion.  Although progenitors of Type II supernovae are also rare, the carbon bands suggest not only that a dC had a companion, but specifically that it had a more massive companion which evolved to produce carbon. Thus, the fraction of dC stars which had a companion massive enough to be a type II supernova progenitor may be larger than would otherwise be expected in a sample of cool dwarf stars.
\\
\indent
In summary, $C_2$ bands in dwarf stars suggest membership in a binary at some point in the stars' history.  Thus, dC stars which have been accelerated to extreme velocity by a mechanism which does not involve passage near the Galactic center are particularly interesting.  Dwarf carbon stars are rare, as are very high velocity stars; it would seem unlikely that multiple objects would share both properties merely by coincidence.\\
\acknowledgements
\indent
We thank C. Rockosi and G. Laughlin for useful discussions, and especially H. Harris for permission to quote unpublished USNO parallax results. We thank the anonymous referee for very useful comments which have improved this manuscript.  PG acknowledges the support of NSF grant AST-1010039.  ECC is supported by an NSF Graduate Research Fellowship. ET acknowledges support from NSF grant AST-1412504. The Figure~2 corner plot was produced using Python \textit{corner}~\citep{dfm}. Funding for the SDSS and SDSS-II has been provided by the Alfred P. Sloan Foundation, the Participating Institutions, the National Science Foundation, the U.S. Department of Energy, the National Aeronautics and Space Administration, the Japanese Monbukagakusho, the Max Planck Society, and the Higher Education Funding Council for England. The SDSS Web Site is http://www.sdss.org/. Some of the data presented herein were obtained at the W.M. Keck Observatory, which is operated as a scientific partnership among the California Institute of Technology, the University of California and the National Aeronautics and Space Administration. The Observatory was made possible by the generous financial support of the W.M. Keck Foundation. The authors wish to recognize and acknowledge the very significant cultural role and reverence that the summit of Mauna Kea has always had within the indigenous Hawaiian community.  We are most fortunate to have the opportunity to conduct observations from this mountain.

{\it Facilities:} \facility{Keck:II (DEIMOS)}, \facility{Sloan}

\clearpage

\clearpage

\clearpage

\clearpage
\end{document}